# Ultraviolet Nanophotonics Enables Autofluorescence Correlation Spectroscopy on Label-Free Proteins With a Single Tryptophan

Prithu Roy,[1] Jean-Benoît Claude,[1] Sunny Tiwari,[1] Aleksandr Barulin,[1] Jérôme Wenger[1,*]

[1] Aix Marseille Univ, CNRS, Centrale Marseille, Institut Fresnel, AMUTech, 13013 Marseille, France

* Corresponding author: jerome.wenger@fresnel.fr

**Abstract:**

Using the ultraviolet autofluorescence of tryptophan aminoacids offers fascinating perspectives to study single proteins without the drawbacks of fluorescence labelling. However, the low autofluorescence signals have so far limited the UV detection to large proteins containing several tens of tryptophan residues. This limit is not compatible with the vast majority of proteins which contain only a few tryptophans. Here we push the sensitivity of label-free ultraviolet fluorescence correlation spectroscopy (UV-FCS) down to the single tryptophan level. Our results show how the combination of nanophotonic plasmonic antennas, antioxidants and background reduction techniques can improve the signal-to-background ratio by over an order of magnitude and enable UV-FCS on thermonuclease proteins with a single tryptophan residue. This sensitivity breakthrough unlocks the applicability of UV-FCS technique to a broad library of label-free proteins.

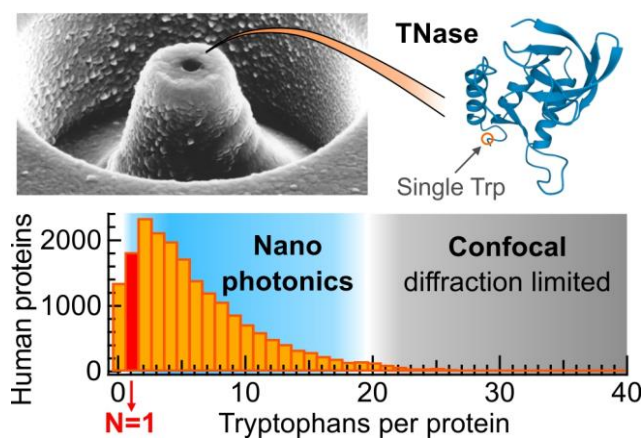

Figure for Table of Contents

Proteins containing aromatic amino-acids (tryptophan, tyrosine and phenylalanine) are naturally fluorescent when excited in the deep ultraviolet (UV).[1–3] This property opens fascinating opportunities to investigate proteins in their native state without introducing any external fluorescent label.[4,5] Avoiding the fluorescence labelling not only simplifies the protein preparation and purification steps, it importantly rules out all the necessary controls to ensure the fluorescent label does not affect the protein.[6–11]

Tryptophan (Trp), the brightest of the three aromatic amino acids, has a 12% quantum yield in water and an absorption cross section 20× lower than typical fluorescent dyes.[1,2] As a consequence, the tryptophan UV autofluorescence signal is very dim as compared to an organic fluorescent dye. Therefore, earlier works on UV (auto)fluorescence correlation spectroscopy (UV-FCS) have been restricted to large proteins featuring several tens of tryptophan residues such as β-galactosidase (156 Trps),[12,13] hemocyanin (148 Trps),[14] phosphofructokinase oligomers (340 Trps) [15] or protein amyloid fibrils (>500 Trps).[16,17] Until now, the smallest proteins detected with UV-FCS are penicillin amidase (29 Trps) and streptavidin (24 Trps).[18] However, tryptophan is among the least abundant amino acids in eukaryotic proteins,[19] and so the vast majority of proteins possess only a few Trp residues. A UniProt database survey out of more than 20,000 human proteins shows that on average a human protein contains about 7 Trp residues, with half of the proteins bearing between 1 and 5 Trps (Fig. 1a and Supporting Information Fig. S1).[20] This shows that in order to really exploit the full potential of UV autofluorescence detection and explore a broad library of label-free proteins, the sensitivity of UV-FCS must be pushed by more than one order of magnitude down to the single tryptophan level.

Here we address this sensitivity challenge and demonstrate UV-FCS on label-free proteins bearing a single Trp residue. This is achieved thanks to the combination of (i) nanophotonic UV antenna to enhance the signal, (ii) detailed analysis to reduce the background intensity and (iii) chemical photostabilizing agents to avoid fluorescence saturation. Our results provide guidelines on how to extend plasmonics into the UV regime [21–30] and further develop label-free single molecule spectroscopy.[4,5,31–33] Earlier works using UV aluminum nanophotonics were restricted to proteins containing a large number of Trp residues such as β-galactosidase (156 Trps) [34,35] and streptavidin (24 Trps).[36] Here we improve the sensitivity by more than one order of magnitude, down to the single tryptophan level. This technical achievement opens the UV-FCS technique to a huge library of proteins bearing only a few Trps (Fig. 1a).

We use a 266 nm deep UV laser excitation and a fluorescence collection in the 310 to 360 nm range for our time-resolved UV confocal microscope (Fig. 1b). To maximize the autofluorescence signal, we employ an optical horn antenna developed recently in our group (Fig. 1c).[35] This nanophotonic device combines a central metal nanoaperture together with a conical horn microreflector. The central nanoaperture of 80 nm diameter concentrates the light in an attoliter detection volume and enhances the autofluorescence from proteins diffusing across this attoliter volume,[34,37] while the metallized conical reflector collects the fluorescence emitted at high angles and steers it towards the microscope objective.[38–40]

Three different proteins are investigated in this work (Fig. 1a): thermonuclease staphylococcal nuclease (TNase) from *Staphylococcus aureus*, transcription antiterminator protein (LicT) from *Bacillus subtilis* and streptavidin from *Streptomyces avidinii* (Strep). TNase is a monomer bearing a single Trp residue, LicT is a homodimer with a total of 2 Trps on the protein dimer, and streptavidin is a homotetramer with a total of 24 Trps. All the details about the proteins used in this work are summarized in the Supporting Information Table S1. TNase was selected because its single Trp residue was theoretically predicted [41] to have a quantum yield of 28% in good agreement with ensemble spectroscopy measurements.[2] TNase is also a widely studied as a model system in protein chemistry.[42,43] LicT was selected for its UV signal being comparable to TNase and its availability in purified form labeled with Cy3B to serve as a control using visible fluorescence spectroscopy.[44,45] Streptavidin was selected because of its higher number of Trps, large availability, moderate mass and good water solubility.

To estimate the feasibility of the UV-FCS detection of proteins with a single tryptophan, we numerically compute the evolution of the FCS correlation amplitude $G(0) = \left(1 - \frac{B}{B+N*CRM}\right)^2 \frac{1}{N}$ as a function of the background intensity $B$, the fluorescence brightness per molecule $CRM$, the number of molecules $N$ and the number of Trp residues per protein (Fig. 1d-f & Fig. S2).[46–48] The ranges of values are taken to reproduce our experiments. For a given number of molecules, the correlation amplitude quickly drops when the signal to background $N\,CRM/B$ decreases owing to the quadratic exponent in the $\left(1 - \frac{B}{B+N*CRM}\right)^2$ term. We indicate on Fig. 1d-f different minimum thresholds for possible FCS detection corresponding to the contours where $G(0)$ amounts to 0.001, 0.005 or 0.01. Since there is no general consensus in FCS in defining this minimum threshold,[46] we decide to show three different values. FCS amplitudes above 0.01 should be easily detectable on a wide range of systems, whereas values below 0.001 are highly challenging as they come close to the electronic noise level and the residual correlation from the background. The calculations results in Fig. 1d and S2 show that maximizing the signal to background ratio $N\,CRM/B$ is crucial to ensure the feasibility of UV-FCS experiments. Using typical values of UV autofluorescence brightness and background intensity representative of our experiments, we compute the predicted correlation amplitudes for the horn antenna (Fig. 1e) and the confocal setup (Fig. 1f) as a function of the number of tryptophan residues per protein, assuming for simplicity that all Trp residues contribute equally to the autofluorescence signal. For the horn antenna, detectable correlation amplitudes above 0.01 are found for a single tryptophan provided the number of proteins in the detection volume is between 2 and 60 (Fig. 1e). UV-FCS on a single tryptophan protein appears feasible with the horn antenna. For the confocal reference, the single tryptophan always yields correlation amplitudes below 0.001 (Fig. 1f). Increasing the number of proteins in the detection volume does not compensate for the lower signal to background ratio in this case. A realistic confocal UV-FCS experiment requires that the protein carries at least 20 Trp residues.

Reducing the background intensity is crucial for improving the sensitivity of UV-FCS (Fig. 1g-i). Figure 1d shows that even with the brightness enhancement brought by the horn antenna, UV-FCS on a single Trp protein would be nearly impossible if the background intensity exceeds 2,000 counts/s. Experiments performed on different samples milled by focused ion beam (FIB) show that the implantation of gallium oxide resulting from the FIB process [49] is the major source of background in the horn antenna (Supporting Information Fig. S3 & S4). Gallium oxide is luminescent when excited in the UV. [50,51] This background contribution can be controlled by FIB while selecting the proper milling depth (Fig. S3). Moreover, the gallium oxide luminescence spectrum is shifted toward 400 nm [50,51] and can be partially separated from the protein autofluorescence spectra with the 310-360 nm bandpass filter (Fig. 1g). The gallium oxide photoluminescence contains a long lifetime component, significantly longer than the 12.5 ns laser repetition period (Fig. S4). Temporal gating to select the photon arrival time in a 3 ns window immediately after the excitation pulse further reduces the background intensity without losing too much of the protein signal (Fig. 1h). The 3 ns window is chosen to correspond to approximately 3× the tryptohan fluorescence lifetime for the different proteins used here in the horn antenna. Altogether, the combination of spectral filtering with temporal gating reduces the background intensity by 7× and improves the signal to background ratio by 2.7× (Fig. 1i) opening the possibility for single Trp detection.

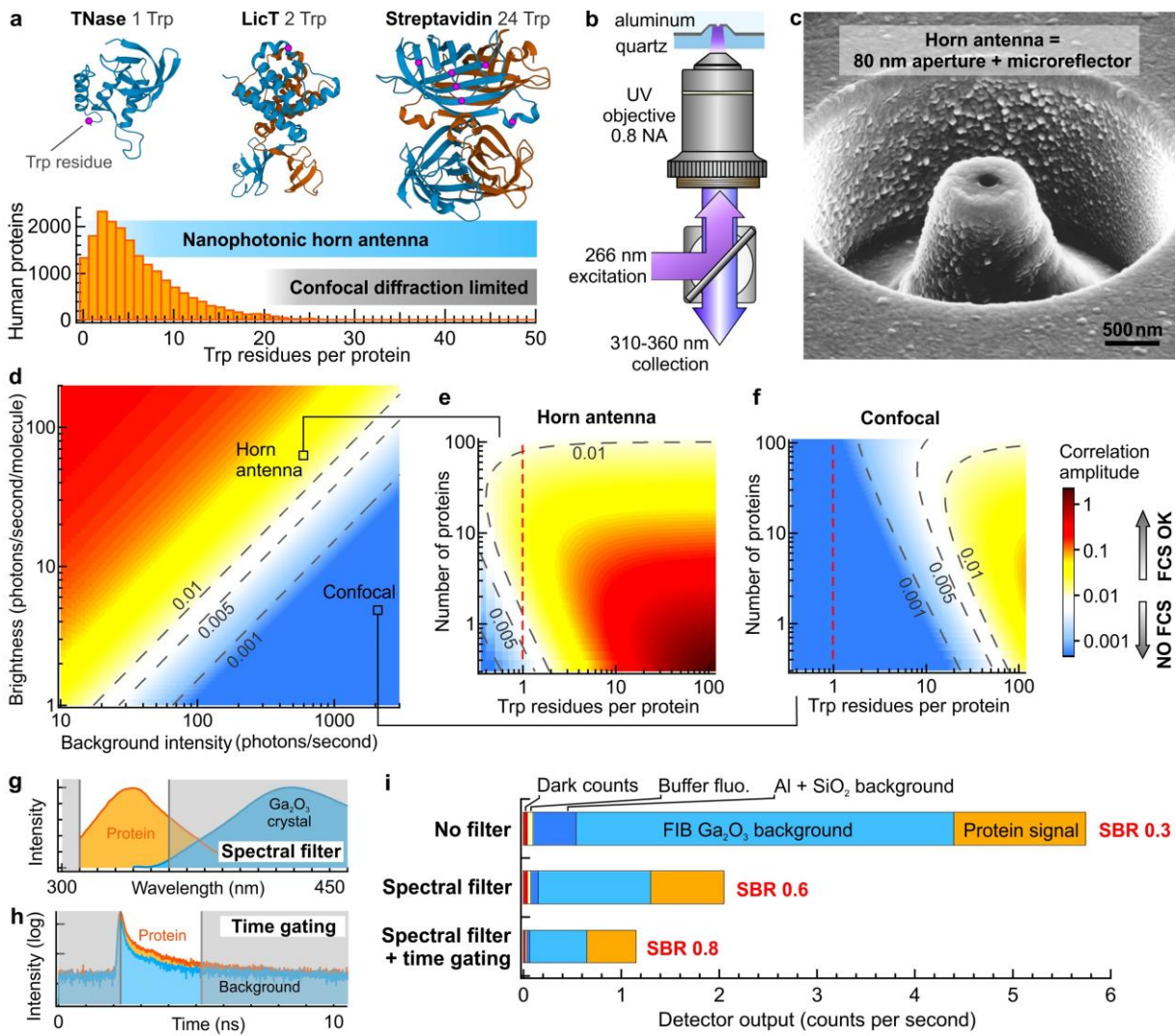

**Figure 1.** Optimizing signal to background ratio to detect a single tryptophan. (a) Histograms of the number of tryptophan residues per protein extracted from a UniProt database of 20 399 reviewed human protein entries.[20] The limits of detection for the confocal [18] and the optical horn antenna (this work) are indicated. The top images show the 3D structures of the proteins used in this work, made using Mol* viewer, with tryptophan residues highlighted by magenta dots on selected monomers.[52] (b) Scheme of the experiment. (c) Scanning electron microscope image of a horn antenna combining a central nanoaperture and a conical reflector. (d) Calculation of the FCS correlation amplitude $G(0)$ as a function of the background intensity $B$ and the fluorescence brightness per molecule $CRM$. A constant number of $N = 5$ proteins was assumed, each protein carrying a single tryptophan. All 2D maps in (d-f) share the same color scale. The lines at 0.001, 0.005 and 0.01 indicate the boundary threshold for possible FCS detection. Correlation amplitudes above 0.01 can be easily detected, while values below 0.001 are highly challenging if not impossible to detect. (e,f) Calculations of the correlation amplitude for the horn antenna and the confocal case as a function of the number of tryptophan residues per protein and the number of diffusing proteins in the detection volume. The values for the background intensity and the fluorescence brightness per tryptophan residue are indicated by the markers in (d) and correspond to typical values in our experiments. (g,h) Strategies to improve the signal to background ratio (SBR) by spectral filtering and time gating. The protein data correspond to 8 µM

TNase solution in the horn antenna (6 proteins in the detection volume). The gallium oxide photoluminescence spectra is taken from ref [50]. (i) Experimental background intensity and total signal (8 µM TNase solution) in the horn antenna upon spectral filtering and time gating.

Along with the reduction of background intensity, another major element determining the UV-FCS sensitivity is the maximization of the fluorescence signal. This involves the use of chemical photostabilizing agents to mitigate the buildup of the radical and triplet state populations leading to fluorescence saturation (Supporting Information Fig. S5).[18,53,54,55] We use mercaptoethylamine (MEA also known as cysteamine) or gluthatione (GSH) which are effective in improving the autofluorescence of both TNase and LicT up to 8× (Fig. S5) without introducing any significant additional background (Fig. S6). GSH is an antioxidant naturally present in the human body to balance oxidative stress and neutralize reactive oxygen species (ROS).[56] For TNase and strepatvidin experiments, we have decided not to use any oxygen scavenging approach as the TNase autofluorescence was not significantly affected by the presence of the oxygen dissolved in the buffer. For LicT experiments, oxygen was removed by bubbling the solution with argon prior to recording the data (see Methods). That way, dissolved oxygen was removed without adding any supplementary background (Fig. S5f-h).

Figure 2 summarizes our main experimental results aimed at pushing the UV-FCS sensitivity down to the single tryptophan level. The linear evolutions of the measured total intensities with the TNase and LicT protein concentrations (Fig. 2a,b) provide a direct control that our experiments are sensitive to the protein autofluorescence signal, even with proteins bearing only one or two Trp residues. The UV-FCS is computed and fitted (Fig. 2c, S7-S8) to extract the number of detected proteins $N$ and their autofluorescence brightness $CRM$ (see Methods). Even in the absence of any protein sample, we still detect a residual background correlation from the horn antenna filled with the buffer solution (gray trace in Fig. 2c). This background correlation appears on all our traces with a long characteristic time above 50 ms and an amplitude below 0.01 (Fig. S7-S9) which may indicate an origin related to some remaining mechanical vibrations or electric noise on our microscope. The significant difference between the characteristic correlation time of this background (> 50 ms) and the protein diffusion time (< 0.5 ms) enables a clear separation of their contributions in the FCS signal so that the contribution from the diffusing proteins can be recovered. Besides, the correlation amplitude related to the protein is always at least twice larger than this residual background correlation (Fig. S7,S8). In the absence of spectral filtering and time gating (Fig. S9), the correlation amplitude found with the TNase protein falls down to 0.002 and cannot be distinguished from the background anymore. Our results for both TNase and LicT are in good agreement with the numerical calculations in Fig. S1d: using the experimentally determined parameters for the background intensity $B$ and the autofluorescence brightness $CRM$, the observed correlation amplitudes follow the theoretical predictions of our model (Fig. S10).

Figure 2d-f compare the statistical distributions of the UV-FCS results for the horn antennas and different proteins. The number of TNase molecules seen by UV-FCS follows a linear dependence with the protein concentration (Fig. 2d). Statistical T-tests confirm the difference between the data distributions. On the contrary, when we probe similar concentrations (8 µM) of different proteins (Fig. 2e), the T-tests give *p*-values above 0.05 and so the number of molecules cannot be clearly distinguished anymore. The fluorescence brightness are different between TNase, streptavidin and LicT (Fig. 2f). However, as reported previously,[18,57–59] the UV autofluorescence brightness does not scale linearly with the number of Trp residues as the presence of nearby aminoacids can quench the Trp emission by charge or energy transfer. Therefore, the brightness for streptavidin is not 24× higher than the brightness for TNase. The average quantum yield for a Trp residue in streptavidin was estimated to be 3.5 ± 1 %,[18,36] while the quantum yield of Trp in TNase was reported to be 28 ± 2 %.[2,41] With these values, the brightness for streptavidin is expected to be 24*3.5/28 = 3.0 ± 0.9 times higher than the TNase brightness. Our experimental results stand in good agreement with this estimate as we find that the streptavidin brightness is 1.9 ± 0.7 times higher as compared to TNase (Fig. 2f). The discrepancy between the expected and the measured ratios is related to the saturation of streptavidin at slightly lower power. We did not monitor any sign of photobleaching in our experiments.

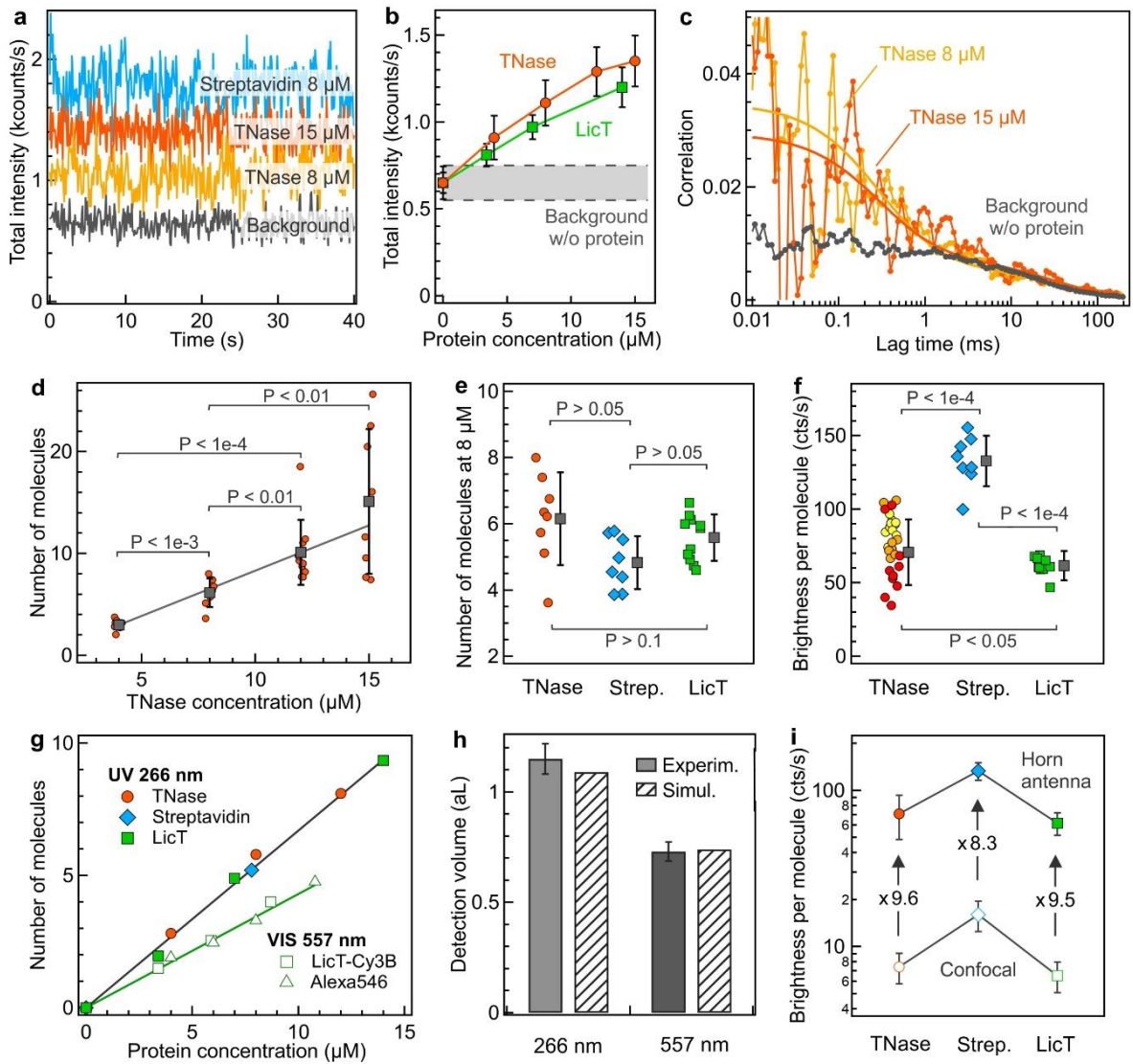

**Figure 2.** Label-free UV-FCS on proteins with a single tryptophan residue enabled by UV horn antennas. (a) Fluorescence time traces and background intensity after spectral filtering and time gating for a single UV horn antenna with different proteins and different concentrations. (b) Fluorescence intensity increase with the protein concentration. The shaded area indicates the background level ± 2 times the standard deviation of the background intensity. (c) FCS correlation traces recorded with the horn antenna for two TNase concentrations. The background correlation is shown in gray, it corresponds to the FCS correlation obtained in the same experimental conditions in the absence of protein target. The fit results are detailed in Supporting Information Fig. S6. (d) Number of TNase proteins determined by FCS in the horn antenna as a function of the TNase concentration. Statistical T-tests have been performed to compare between the distributions in (d-f), the resulting $p$-values are written on the graphs for each pair of distributions. In (d-f), color markers represent individual measurements, gray squares represent the average ± one standard deviation. (e) Comparison between the numbers of proteins detected by FCS for three different proteins at the same 8 μM concentration. (f) Fluorescence brightness per molecule $CRM$ determined by FCS for the different proteins. (g) Average number of molecules measured by FCS in the horn antenna as a function of the concentration for the three different proteins in the UV (color markers, the black line is a fit). The error bars on the individual

data are similar to the ones in (d) and are not represented for clarity. Empty markers show the numbers of molecules determined by FCS using Alexa 546 and Cy3B fluorescent dyes at 557 nm excitation. The slope of the lines are proportional to the detection volume, which is shown in (h) for both UV (266 nm) and visible (557 nm) laser excitation. Numerical simulations of the detection volume (patterned area) confirm the FCS results. (i) Fluorescence brightness per protein in the horn antenna as compared to the confocal reference for the three different proteins. The data points for the horn antenna correspond to the average $CRM$ determined by FCS. To determine the number of molecules for the confocal data, we assume a 1.8 fL confocal volume on our UV microscope.

The UV-FCS average number of molecules is plotted in Fig. 2g as a function of the protein concentration for the different proteins used in this work (filled markers in Fig. 2g). We find that the different datasets follow the same line whose slope is proportional to the size of the detection volume. Experimentally, we determine a detection volume of 1.15 ± 0.07 attoliter ($10^{-18}$ L), in good agreement with the numerical simulations (Fig. 2h). To further validate the UV-FCS data, the same 80 nm apertures in aluminum are probed with visible fluorescent dyes and 557 nm laser excitation.[60] We use a Cy3B label on LicT protein and the free fluorescent dye Alexa Fluor 546 to perform visible FCS experiments recording the number of fluorescent molecules as a function of the concentration (raw FCS data are shown in Fig. S11). The data for both LicT-Cy3B and Alexa Fluor 546 follow the same linear relationship with the concentration defining a similar detection volume inside the aperture (Fig. 2g). Because of the longer illumination wavelength (557 instead of 266 nm) the penetration depth inside the nanoaperture [61] and hence the size of the detection volume are different between the UV and the visible experiments. This difference can be accounted for by the numerical simulations (Fig. 2h & S12), providing a supplementary control of the experimental results.

We compare the brightness between the three different proteins using the horn antenna and the confocal reference (the brightness for the confocal case is estimated from the measured concentration and the 1.8 fL value of the confocal volume). The horn antenna improves the brightness by about 9× for all the different proteins (Fig. 2i). As the horn antenna is a weakly resonant structure (as compared to a dimer nanogap antenna[62,63]), its gain is essentially brought by the local excitation intensity increase and the improved collection efficiency.[35] The quantum yield enhancement plays a minor role here,[36] so similar net fluorescence enhancement are expected despite proteins with different Trp quantum yields are used. The 9× enhancement stands also in good agreement with our calibration using the UV fluorescent dye p-terphenyl.[35] If we compare between the best results found for the horn antenna (single Trp brightness 70 cts/s, background 600 cts/s) and the confocal setup (single Trp 8 cts/s, background 2000 cts/s), our combined solution improves the signal to background ratio by 30×. The 1000× lower detection volume with the antenna efficiently eliminates the background intensity stemming from the solution (Fig. S6), yet at the expense of a supplementary background from the antenna luminescence (Fig. 1i).

In presence of background, the signal to noise ratio for determining $G(0)$ is given by $SNR = CRM \left( \frac{SBR}{1+SBR} \right) \sqrt{T_{tot} \, \Delta\tau}$ (see Supporting Information section S14), where $SBR = (N \, CRM)/B$ is the signal to background ratio, $T_{tot}$ is the total integration time and $\Delta\tau$ is the temporal width of the counting interval. The $SNR$ provides a figure of merit to compare between experiments and discuss the feasibility of an FCS experiment. For the horn antennas and the proteins used here, the $SNR$ ranges from 0.7 to 1.5 (Fig. S7,S8), while for the confocal configuration and even with one hour integration time, the $SNR$ remains below 0.2. This consideration further highlights the key role played by the nanophotonic antenna to enable UV-FCS on proteins featuring a low number of Trp residues. Besides, the FCS diffusion time influences the choice of the temporal width $\Delta\tau$, as $\Delta\tau$ must remain significantly smaller than the FCS diffusion time. The detection of slower diffusing species enables the use of a longer counting interval $\Delta\tau$ which improves the signal to noise ratio and can partly compensate for a lower autofluorescence brightness.

As most proteins bear only a few tryptophan residues, being able to detect a single tryptophan (instead of several tens) is a major breakthrough opening the possibility to apply the UV-FCS technique to a huge library of label-free proteins. This challenging task requires a careful optimization of the signal to background ratio combining approaches to maximize the signal (optical horn antenna, antioxidants) and reduce the background intensity (FIB milling depth, spectral filtering, time gating, buffer composition). Our calculations provide useful guidelines to predict the feasibility of the experiments based on the correlation amplitude and the signal to noise ratio. Altogether, the data presented in Fig. 2 demonstrate that a protein bearing a single Trp residue can be detected using UV-FCS. We envision that the methods developed here to optimize the UV autofluorescence signal to background ratio will be useful to a wide range of future studies on label-free single protein spectroscopy,[4,5,31–33] as well as the advancement of plasmonics into the UV range.[21,22,25,26] UV-FCS can provide information about local concentration, diffusion properties, and autofluorescence brightness per molecule to shine new light on protein interaction dynamics with ligands or other molecular partners.[46–48] While in scattering microscopy the interference signal scales with the 3rd power of the nanoparticle diameter,[4,33,64,65] UV-FCS is less sensitive to the protein size, relying more its tryptophan content. The TNase proteins detected here have a molecular weight lesser than 20 kDa, opening the possibility to detect label-free proteins with molecular weights in the single-digit kDa range. As supplementary advantage of the technique, the detection volume is in the attoliter range, three orders of magnitude below that of a diffraction-limited confocal microscope, so that single molecule detection and UV-FCS can operate at micromolar concentrations.[66,67] Being able to work at high concentrations with single molecule resolution and/or FCS is essential to study a broad range of enzymatic reactions, protein-protein and protein-DNA/RNA interactions with Michaelis constants or dissociation constants in the micromolar range.[68,69]

**Supporting Information**

Tryptophan occurrence in human proteins, Protein information and sequences, Calculations of FCS correlation amplitude in presence of background, Background intensity dependence with the FIB milling depth, Background reduction using spectral filtering, Autofluorescence signal improvement using antioxidants, Background from the buffer solution in the confocal configuration, FCS correlation traces and fit results, FCS correlation is observed in the absence of spectral filtering and time gating, Validation of the experimental correlations with the calculations model, Control FCS experiments on visible fluorescent dyes, Numerical simulations of the detection volume, Protein absorption and emission spectra, Signal to noise ratio in FCS in presence of background, Supplementary methods.

**Data availability**

All data are available from the corresponding author upon request.

**Note**

The authors declare no competing interest.

**Author contributions**

J.W. designed and supervised research; P.R. performed research and analyzed data; A.B. built the microscope and contributed to preliminary experiments; J.B.C. fabricated horn antennas; S.T. and P.R. performed electromagnetic simulations; J.W. wrote the paper.

**Acknowledgments**

We thank Emmanuel Margeat, Nathalie Declerck and Caroline Clerté for providing the LicT proteins. This project has received funding from the European Research Council (ERC) under the European Union's Horizon 2020 research and innovation programme (grant agreement No 723241).

**Supporting Information for**

**Ultraviolet Nanophotonics Enables Autofluorescence Correlation Spectroscopy on Label-Free Proteins With a Single Tryptophan**

Prithu Roy,[1] Jean-Benoît Claude,[1] Sunny Tiwari,[1] Aleksandr Barulin,[1] Jérôme Wenger[1],*

[1] *Aix Marseille Univ, CNRS, Centrale Marseille, Institut Fresnel, AMUTech, 13013 Marseille, France*

*\* Corresponding author: jerome.wenger@fresnel.fr*

**Contents:**

## S1. Tryptophan occurrence in human proteins

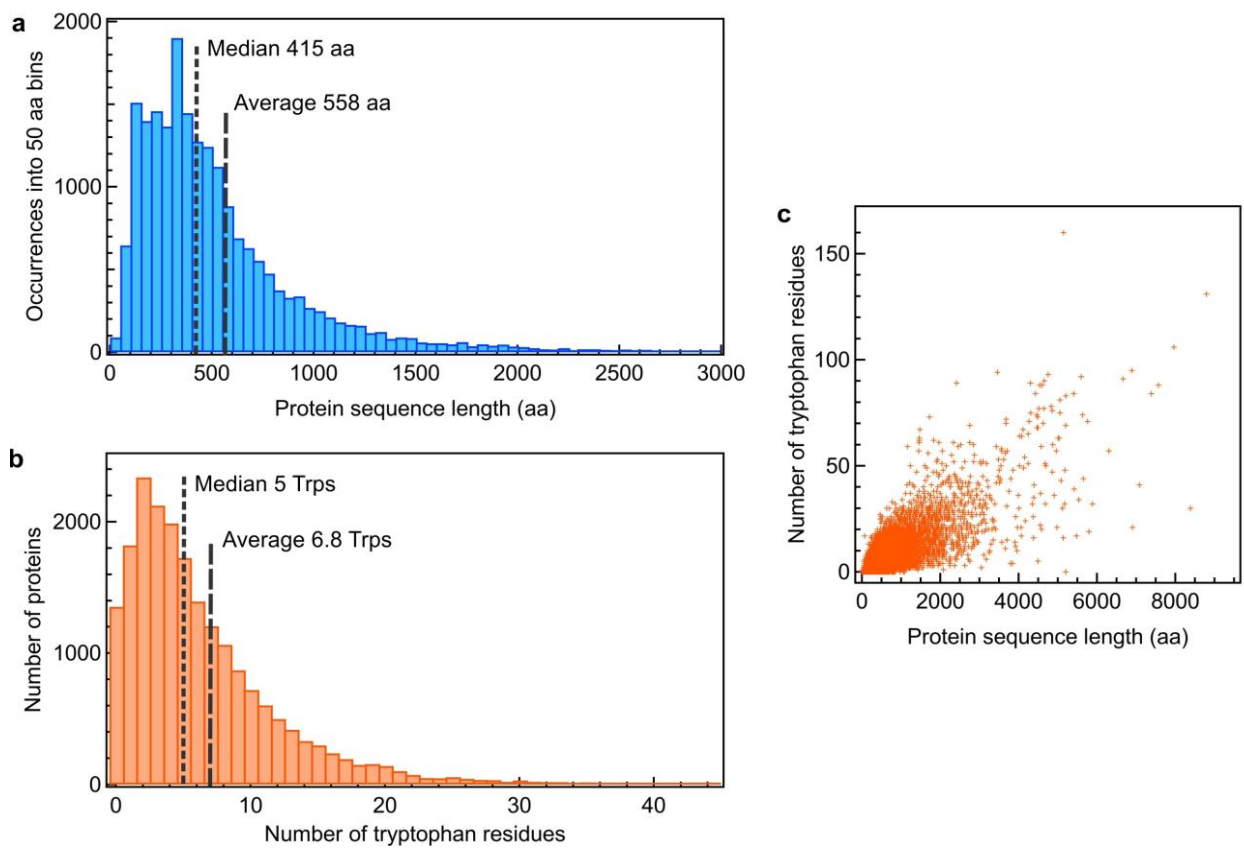

**Figure S1.** Statistics of tryptophan distribution in human proteins. A total of 20 399 proteins was extracted from the UniProt database corresponding to human proteins whose entries were reviewed by Swiss-Prot [1]. From the sequence information, the histograms of protein sequence length (a) and number of tryptophan residues (b) are extracted. The scatter plot in (c) shows the correlation between the number of Trp residues and the total sequence length and enables a better view of the extreme values of the distributions. Out of the 20 399 human proteins with available data, 1353 do not bear a Trp residue (6.6% of total), so 93.4% of humans proteins have at least one Trp residue. The Trp frequency measured from the entire distribution is 1.2%, which corresponds also to the ratio of the average values (on average, a human protein contains 558 amino acids among which 6.8 (1.2%) are tryptophans). About half of the human proteins (48.9%) have between 1 and 5 tryptophans. Only 4% of the human proteins have more than 20 Trp residues.

**Table S1.** Detailed information about the proteins used in this work.

| Acronym | TNase | Strep | LicT |
|---|---|---|---|
| Name | Thermonuclease Staphylococcal nuclease | Streptavidin | Transcription antiterminator protein LicT |
| Organism | Staphylococcus aureus | Streptomyces avidinii | Bacillus subtilis |
| UniPROT reference | P00644 | P22629 | P39805 |
| RCSB PDB structure ref | 1A2U | 2RTR | 6TWR |
| Sigma Aldrich product number | N 3755 | S 4762 | n/a |
| Form | Monomer | Homotetramer | Homodimer |
| Monomer molecular weight (Da) | 16,807 | 18,834 | 19,908 |
| Monomer sequence length (aa) | 149 | 183 | 175 |
| Monomer sequence (tryptophan **W** and tyrosine **Y** highlighted) | ATSTKKLHKEPATLIKAIDGDTVKLM**Y**KGPQMTFRLLLVDTPQTKHPKKGVEK**Y**GPEASAFTKKMVENAKKIEVEFNKGQRTDK**Y**GRGLA**Y**I**Y**ADGKMVNEALVRQGLAKVA**Y**V**Y**KPNNTHEQLLRKEKKSEAQAKLNI**W**SENDADSGQ | MRKIVVAAIAVSLTTVSITASASADPSKDSKAQVSAAEAGITGT**W**Y**NQLGSTFIVTAGADGALTGT**Y**ESAVGNAESR**Y**VLTGR**Y**DSAPATDGSGTALG**W**TVA**W**KNN**Y**RNAHSATT**W**SGQ**Y**VGGAEARINTQ**W**LLTSGTTEANA**W**KSTLVGHDTFTKVKPSAASIDAAKKAGVNNGNPLDAVQQ | MKIAKVINNNVISVVNEQGKELVVMGRGLAFQKKSGDDVDEARIEKVFTLDNKDVSEKFKTLL**Y**DIPIECMEVSEEIIS**Y**AKLQLGKKLNDSI**Y**VSLTDHINFAIQRNQKGLDIKNALL**W**ETKRL**Y**KDEFAIGKEALVMVKNKTGVSLPEDEAGFIALHIVNAELNELQHHHHHH |
| Tryptophan count per monomer | 1 | 6 | 1 |
| Tyrosine count per monomer | 7 | 6 | 4 |
| Extinction coefficient $\varepsilon$ at 280 nm ($M^{-1}$ $cm^{-1}$) | 22,050 | 169,360 | 12,785 |

## S3. Calculations results show that the background intensity plays a major role in determining the feasibility of UV-FCS experiments on label-free proteins

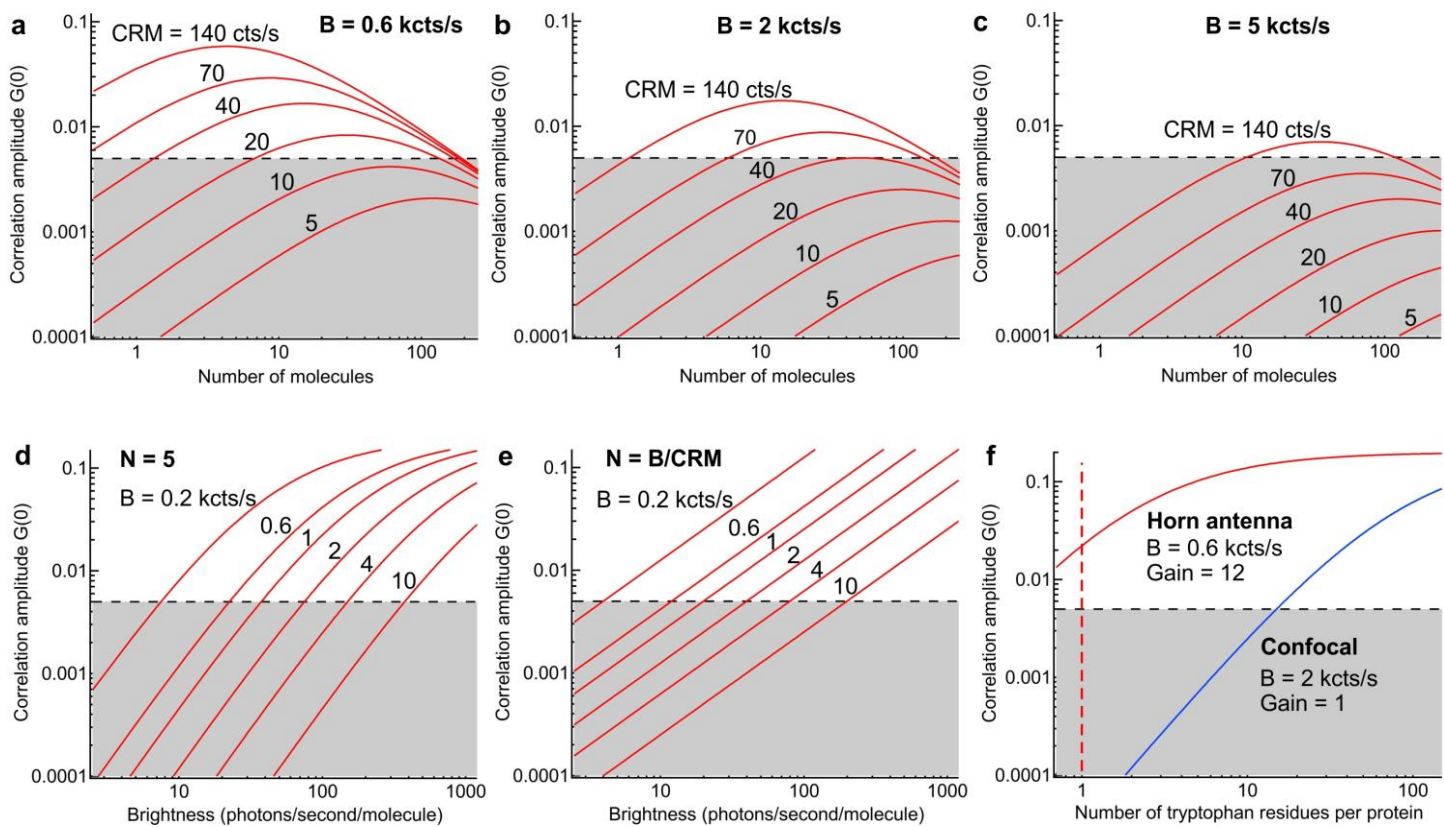

**Figure S2.** Calculations of the FCS correlation amplitude $G(0) = \left(1 - \frac{B}{F}\right)^2 \frac{1}{N}$, where $B$ is the background intensity, $N$ the number of detected fluorescent molecules, and $F = B + N * CRM$ is the total detected intensity in presence of the fluorescent molecules. $CRM$ denotes the average fluorescence brightness per molecule (count rate per molecule). (a-c) Correlation amplitude as a function of the number of detected molecules $N$ for different values of the fluorescence brightness per molecule $CRM$ and background intensity $B$ (as indicated on the graphs, these values are typical for the experiments performed here). The shaded area indicates correlation amplitudes below 0.005 which are highly challenging to detect with FCS due to the residual background correlation and the electronic noise. We define the threshold of G(0) = 0.005 as a minimum amplitude for possible detection. This definition is an arbitrary choice, any value in the range 0.001-0.01 would be realistic. As a consequence of the low signal to background ratio, the $G(0)$ dependence with $N$ is more complex than just the classical $1/N$ rule. For low $N$ values, the background term $\left(1 - \frac{B}{F}\right)^2 = \left(1 - \frac{B}{B+N*CRM}\right)^2$ plays a major role and the correlation amplitude increases when the number of molecules grows. For large $N$ values, the background term is not so influential, and we retrieve the $1/N$ dependence leading to a decrease of $G(0)$ when $N$ grows. It can be shown that the correlation amplitude $G(0)$ reaches its maximum when the number of molecules amounts to $N = B/CRM$. In that case, the maximum amplitude is $G_{max}(0) = CRM/4B$. (d) Correlation amplitude as a function of the $CRM$ for different values of background $B$. For simplicity, the number of molecules is fixed to $N = 5$. (e) Maximum $G_{max}(0)$ value at the optimum number of molecules $N = B/CRM$ corresponding to the highest correlation amplitude achievable for a given $(B, CRM)$ set of values. The calculations results throughout (a-e) show that the background intensity plays a major role in determining the feasibility of an FCS experiment at a given brightness $CRM$. For the range of brightness between 10 and 100 counts/s/molecule typically achievable in our experiments,

the background intensity must remain below 2-3 kcts/s to yield a detectable FCS correlation. (f) Simulated correlation amplitudes as a function of the number of tryptophan residues for the cases corresponding to the horn antenna (red trace) and the confocal reference (blue). The total number of protein is set to $N = 5$ and we assume a constant brightness of 5 counts per second for each tryptophan residue, which is typical for our UV microscope. We also assume a gain of 12 brought by the presence of the horn antenna, which enhances the tryptophan brightness to 12*5 = 60 counts per second. The graph in (f) shows that the detection of a protein with a single tryptophan is feasible with the horn antenna, while for the confocal case, at least 15 tryptophan residues per protein are needed to yield a detectable FCS correlation amplitude.

**S4. Background intensity dependence with the FIB milling depth indicate that the gallium oxide implanted during FIB milling has a major contribution in the total background intensity**

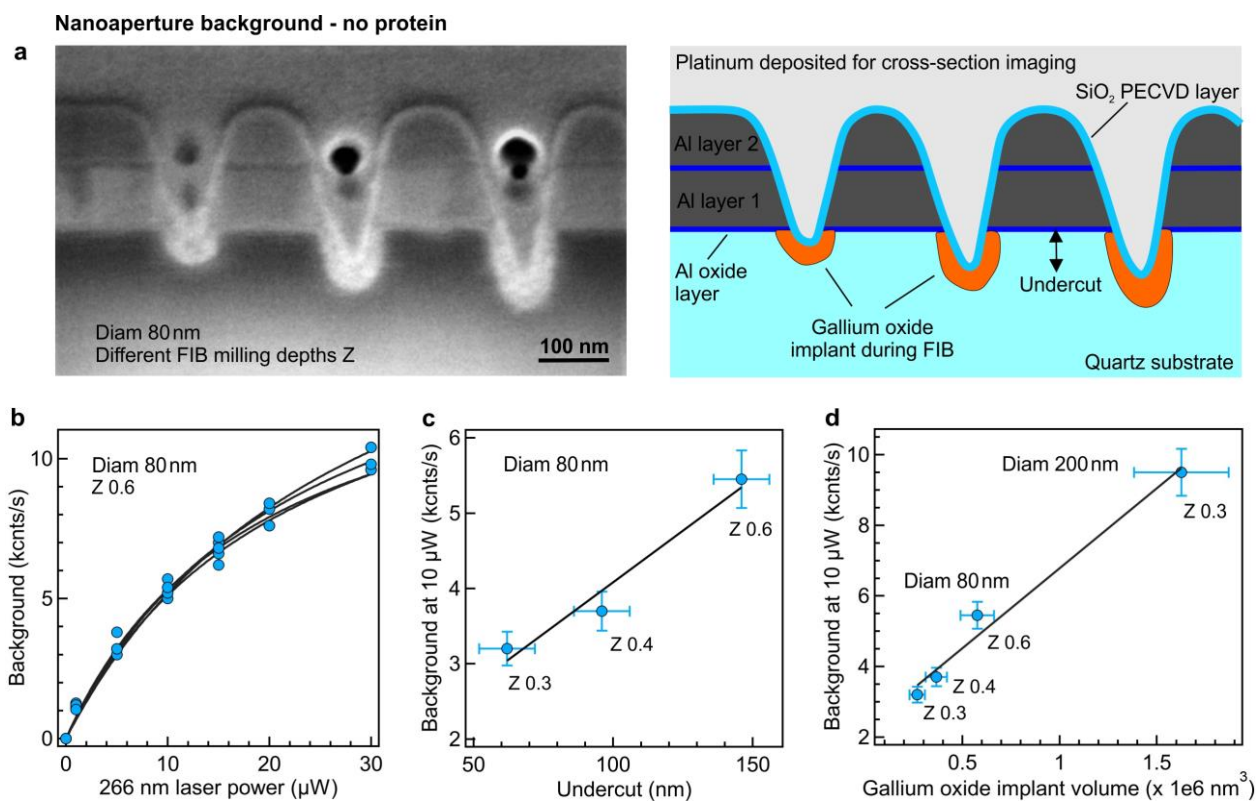

**Figure S3.** (a) Cross-cut scanning electron microscope image of nanoapertures milled with different depths Z. The sample has been filled with platinum for a better side-view imaging, cut by FIB up to the half of the aperture, and tilted by 52° to enable a cross-cut view of the nanoaperture profile. The scheme on the right is a guide to understand the SEM image. Due to the FIB milling process, the region at the bottom of the nanoaperture is enriched in gallium, forming different alloys with the $SiO_2$ quartz substrate, notably gallium oxide $Ga_2O_3$ which is UV-photoluminescent [2,3]. The aperture on the middle corresponds to the configuration used for milling the central aperture of the horn antenna in this work, using a 60 nm deep undercut into the quartz substrate to maximize the signal enhancement [4]. The aperture on the left is not completely milled, leading to a degradation of the signal. (b) Background intensity from a single nanoaperture as a function of the 266 nm laser power. The different data points correspond to different nanoapertures milled with the same conditions. The black lines are numerical fits using a fluorescence saturation model $A*P_{laser}/(1+P_{laser}/P_{sat})$. The saturation indicates that the background stems mostly from photoluminescence, as scattering and backreflection scale linearly with the excitation power. (c) Evolution of the background intensity with the undercut depth for single apertures milled with different Z parameter conditions (the Z parameter is the input used by our FEI DB235 focused ion beam system). (d) Assuming a uniform thickness of 20 nm for the FIB gallium oxide implant, we compute the total volume of gallium oxide implant and plot the background intensity as a function of this volume for 80 nm apertures and 200 nm apertures with different milling depths. The linear relationship between the total background intensity and the volume of the gallium implanted region confirms that the gallium implantation during FIB is the main source of background in our system. We have tried to further reduce this background by annealing to 400°C or photobleaching with prolonged illumination with UV light but both were unsuccessful.

## S5. Background reduction using spectral filtering

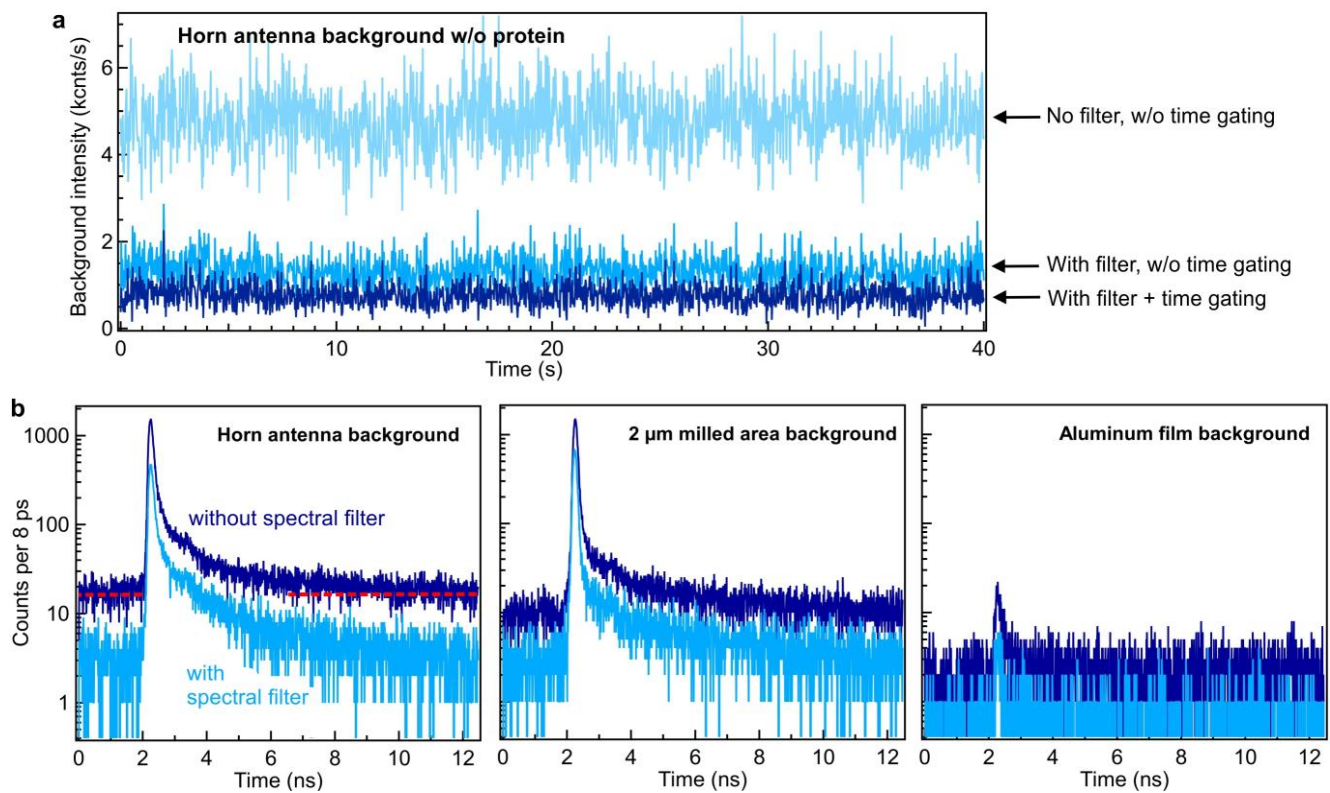

**Figure S4.** (a) Background intensity recorded on a single horn antenna with 10 µW average excitation power with and without spectral filtering and time gating. (b) Time-resolved TCSPC photon arrival time histograms respective to the 80 MHz synchronization signal from the pulsed laser. All histograms are sampled into 8 ps time bins and are integrated over the same 20 s duration so that the intensities on the left axis can be directly compared across the different curves. Dark blue traces correspond to the background without any spectral filter while light blue traces are with the 310-360 nm bandpass filter.

The horn antenna background intensity significantly differs from the background recorded on the bare aluminum film. This shows that the backreflection from the laser and the residual luminescence from the quartz substrate and the aluminum layer are not the primary source of background. On the contrary, the horn antenna background resembles very closely to the background recorded in the center of a 2 µm diameter aperture (no aluminum is illuminated in this case, only the photoluminescent gallium oxide implanted during FIB remains. We used pure heavy water to reduce the background from the solution in this experiment).

Without the spectral filter, the horn antenna background TCSPC histogram has a baseline at around 15 counts per 8 ps binning (corresponding to 1.2 kcnts/s average intensity). This baseline is representative of a photoluminescence with a lifetime much larger than the 12.5 ns period of the laser pulses, so that the arrival time of the photoluminescence photons appears uncorrelated with the laser pulses synchronization signal. The spectral filter strongly reduces this baseline, indicating that most of the baseline photoluminescence falls outside the 310-360 nm region (as expected for gallium oxide photoluminescence[2]). A similar baseline is also found on the 2 µm wide FIB-processed region where the aluminum layer has been totally removed and gallium oxide has been implanted.

## S6. Autofluorescence signal improvement using antioxidants

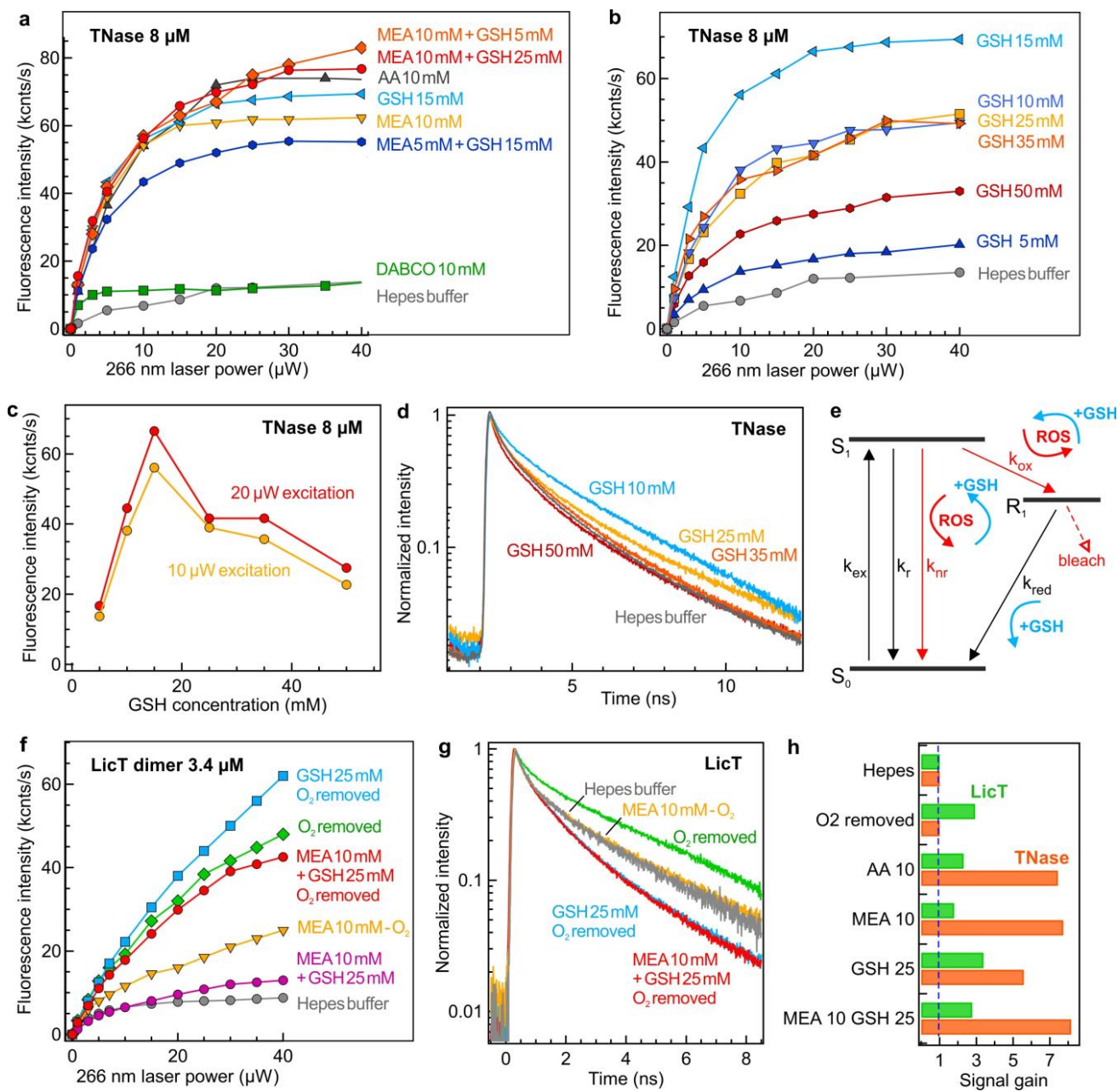

**Figure S5.** (a) Total fluorescence intensity collected from the confocal volume as a function of the 266 nm laser power for a 8 μM TNase solution in presence of different chemical agents (MEA mercaptoethylamine or cysteamine; GSH glutathione; AA ascorbic acid; DABCO 1,4-diazabicyclo[2.2.2]octane). Oxygen removal using degassing did not modify the results for TNase significantly, therefore the oxygen dissolved in the buffer solution was not modified for the TNase experiments. (b,c) Evolution of the TNase fluorescence intensity for different GSH concentrations and the associated fluorescence lifetime decays (d). Moderate GSH concentrations in the 5-25 mM range improve the signal intensity by increasing the quantum yield, the fluorescence lifetime and the excitation power leading to saturation. This indicates a reduction of the nonradiative decay rate as well as a reduction of the radical state buildup [5,6]. However, higher GSH concentrations lead to a decrease of the signal, which may indicate the formation of GSH-TNase complex and/or quenching of the tryptophan singlet excited state by concentrated GSH [7]. (e) Simplified Jablonski diagram of the ground state $S_0$, the singlet excited state $S_1$ and the radical state $R_1$ (the triplet state is omitted for simplicity). The presence of reactive oxygen species ROS produced by UV illumination [6,8] promote

nonradiative decay and oxidation rates. The addition of antioxidant GSH or MEA neutralize the negative effects of ROS and can also promote the reduction from $R_1$ to $S_0$ [6]. However, high concentrations of GSH above 25 mM also tend to increase the nonradiative decay $k_{nr}$ and quench fluorescence. (f,g) Same as (a,d) for LicT proteins. Here, the removal of oxygen using argon degassing plays a beneficial role, improving the signal linearity and delaying the occurrence of saturation. Oxygen removal also tends to increase the fluorescence lifetime and promote the quantum yield, yet the addition of reductants (MEA, GSH) is further reducing the fluorescence lifetime without significantly changing the total detected intensity. (h) Signal gain at 10 μW power for TNase and LicT in presence of different antifading compositions as compared to the hepes buffer reference. For LicT, oxygen is removed by argon degassing while for TNase we keep the oxygen dissolved in the buffer as this did not modify significantly our results. The FCS experiments with the horn antennas use 10 mM MEA and 25 mM GSH with and without oxygen for TNase and LicT respectively.

Our data in Fig. S5 show that moderate MEA and GSH concentrations in the 5-25 mM range improve the linearity of the autofluorescence intensity with the excitation power and increase the autofluorescence intensity and the lifetime. These features correspond to a reduction of the nonradiative decay rate as well as a reduction of the radical state buildup, which we relate to a neutralization of ROS by the antioxidants. We have also found that the oxygen scavenger system GODCAT (glucose oxidase and catalase enzymes) was leading to a too high background for our goal and was prone to photopolymerization issues in the nanoaperture.

## S7. Background from the buffer solution in the confocal configuration

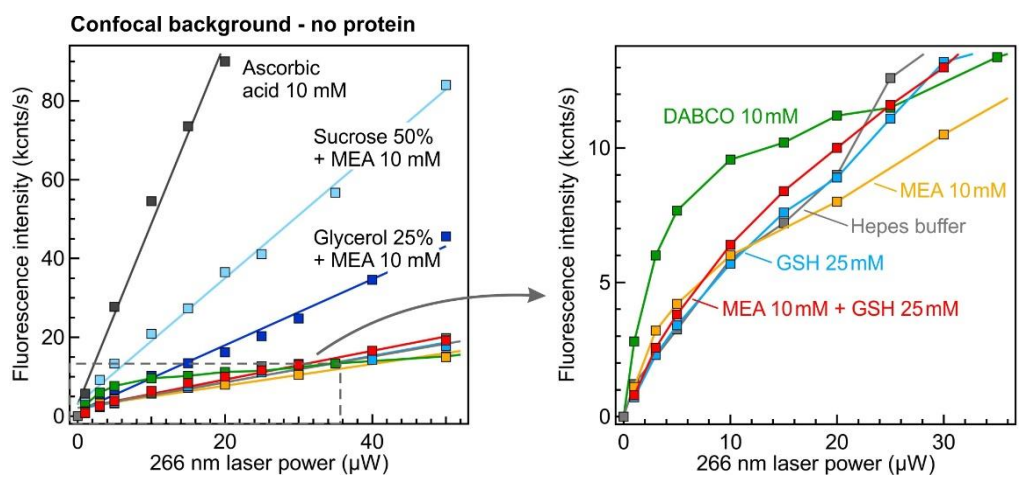

**Figure S6.** Background intensity for the confocal microscope (without horn antenna). Ascorbic acid yields a high background intensity and is discarded as reducing agent. MEA and GSH do not show a significantly higher background intensity than the normal buffer. Impurities present in sucrose and glycerol contribute to increase significantly the background, therefore we have discarded their use to increase the buffer viscosity.

## S8. FCS correlation traces and fit results

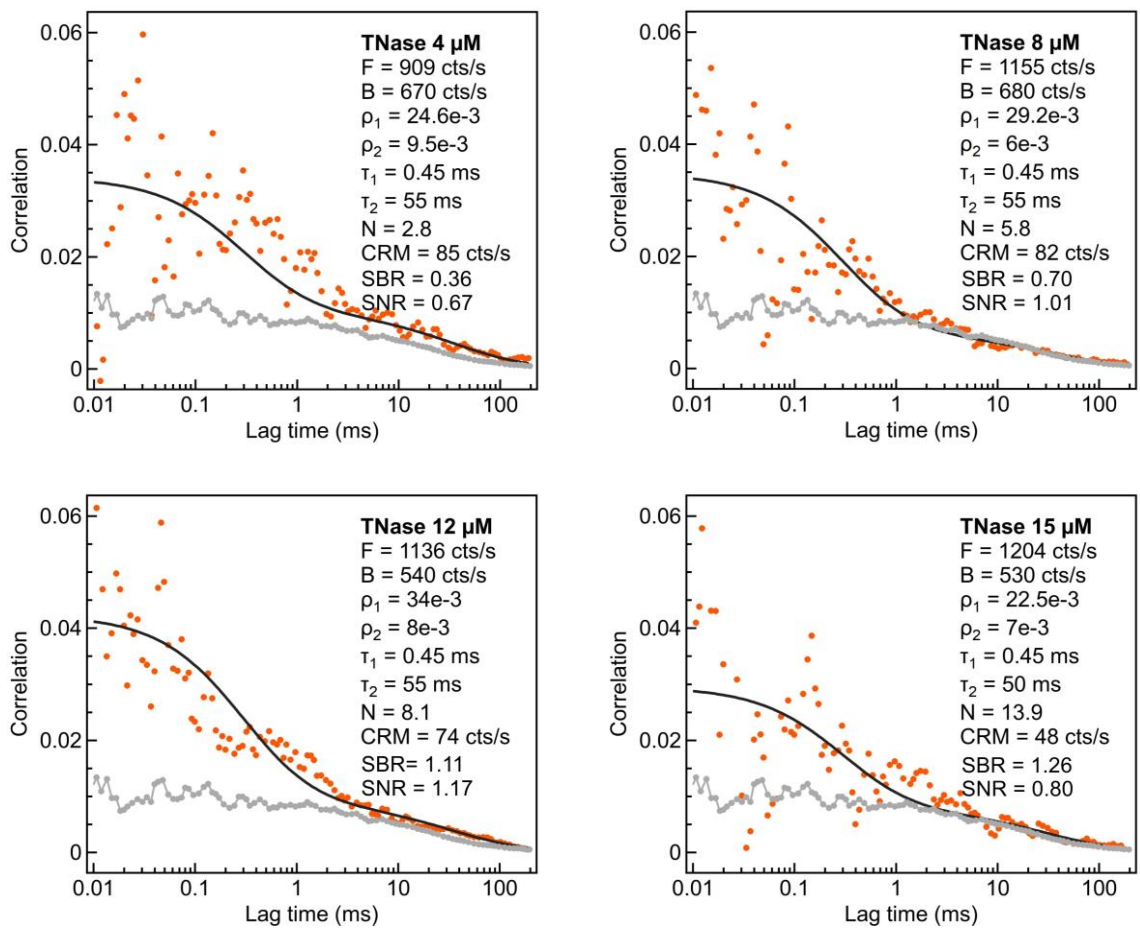

**Figure S7.** FCS correlation functions for different TNase concentrations in the horn antenna (orange dots) and their numerical fits (black line) with spectral filtering and time gating. The fit parameters are indicated for each case. We also indicate the values for the signal to background ($SBR$) and the signal to noise ($SNR$). For definitions of these quantities, see section 14 page S16 of this document. The gray data trace is the residual background correlation recorded in the absence of proteins.

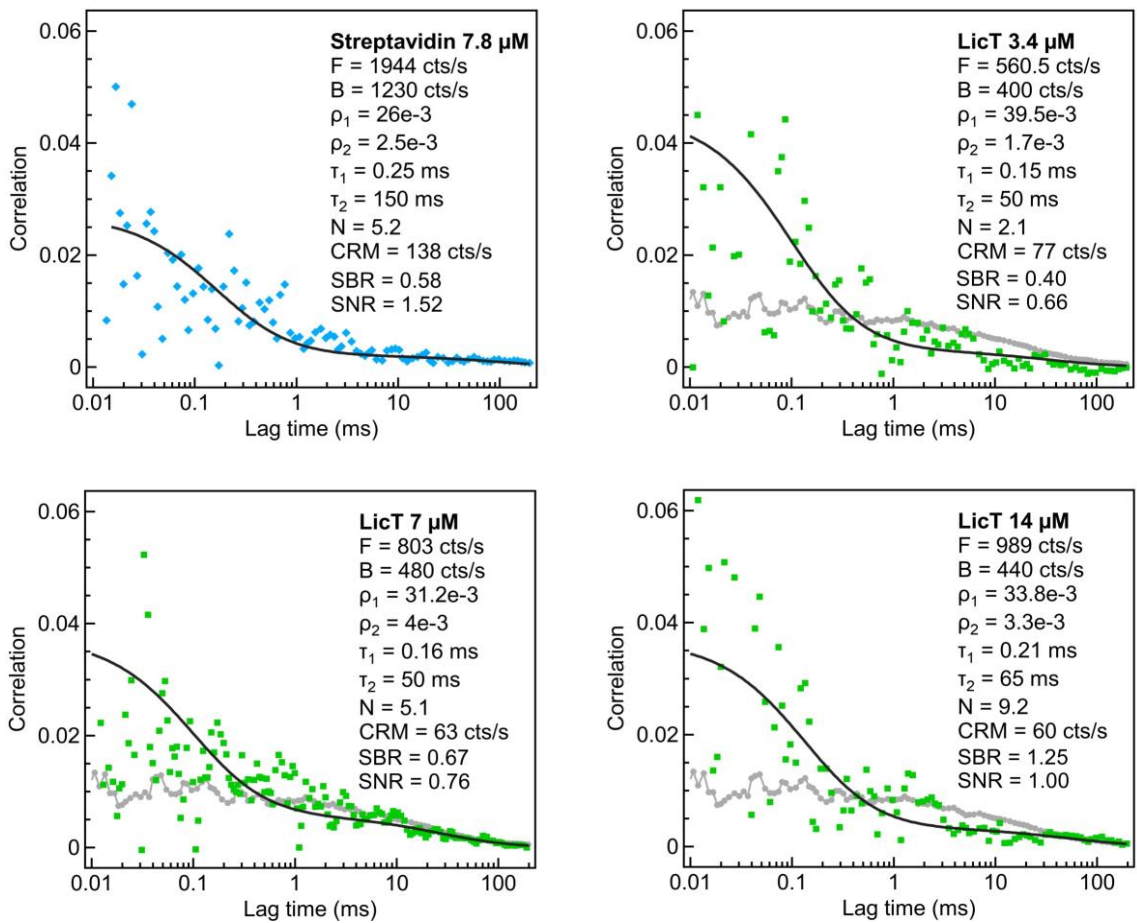

**Figure S8.** FCS correlation functions for streptavidin and LicT in the horn antenna (blue and green markers) and their numerical fits (black line) with spectral filtering and time gating. The fit parameters are indicated for each case as well as the corresponding signal to background ($SBR$) and signal to noise ($SNR$). The gray data trace is the residual background correlation recorded in the absence of proteins.

## S9. No FCS correlation is observed in the absence of spectral filtering and time gating

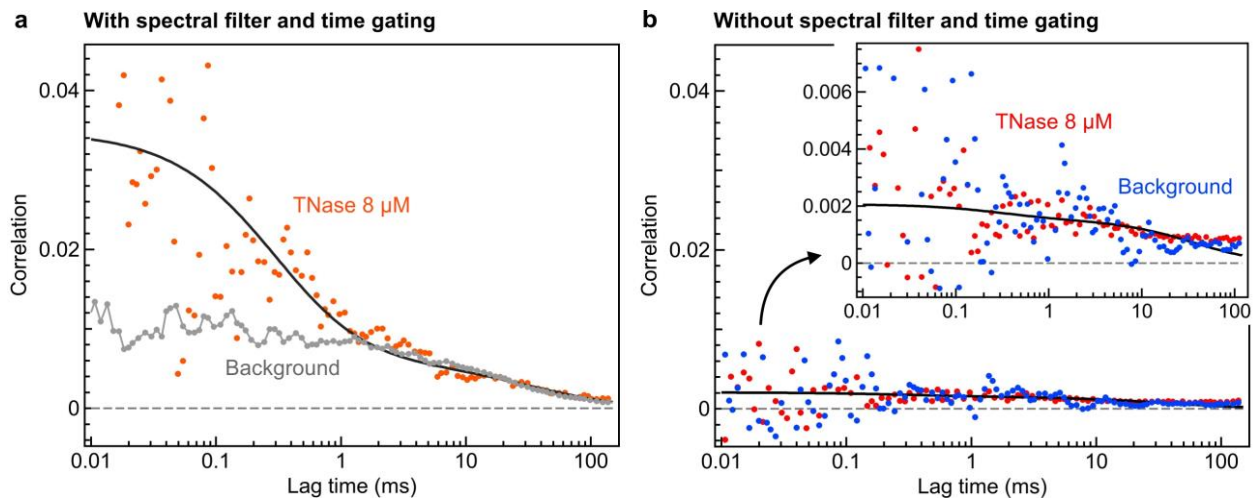

**Figure S9.** Comparison of the FCS correlation with (a) and without (b) spectral filtering and time gating. The horn antenna and TNase concentration are identical for both cases. Due to the higher background level in the absence of spectral filtering and time gating, the correlation amplitude is strongly decreased and becomes undistinguishable from the background correlation. From the calculations in Fig. S1d, for a background intensity of 4.4 kcnts/s, 5 proteins and a brightness of 70 cnts/s, the estimated G(0) amplitude is about 0.001, in good agreement with the experimental results in (b).

## S10. Validation of the experimental correlations with the calculations model

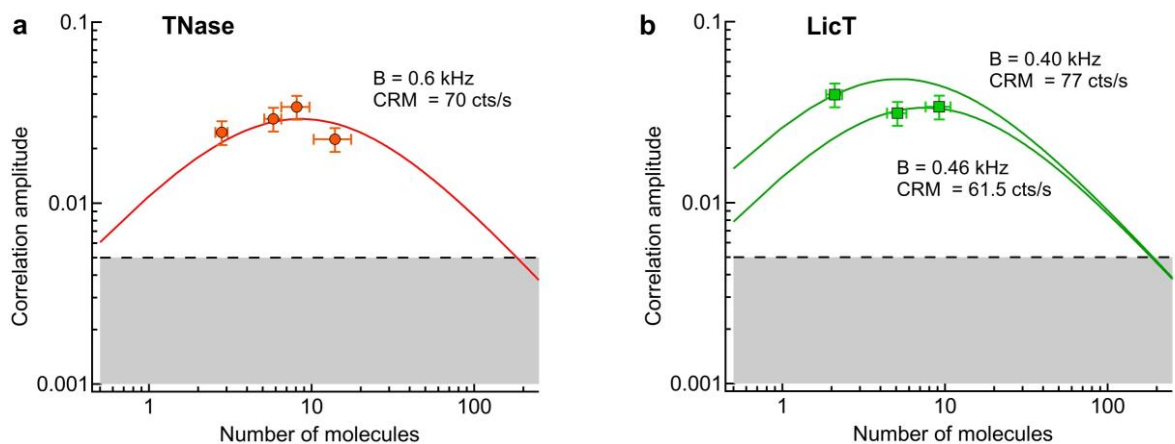

**Figure S10.** Comparison of the experimental correlation amplitudes for the first species $\rho_1$ (markers, corresponding to proteins contribution separated from background) with the theoretical predictions (lines) from Fig. S1 for TNase (a) and LicT (b). For LicT, we used a specially selected sample with low background (presumably because of a slightly reduced milling depth during the FIB process, the background values were recorded for each structure). Two lines are shown for LicT corresponding to slightly different parameters for the background intensity $B$ and the fluorescence brightness $CRM$. The dispersion remains within the uncertainty margins on our experimental data. The shaded region corresponds to the detection limit as in Fig. S2.

## S11. Control FCS experiments on visible fluorescent dyes

To confirm the UV FCS experiments, we have performed FCS measurements on a visible fluorescence microscope using Alexa 546 and Cy3B fluorescent dyes. The experimental setup is described in details in ref. [9] and features a 557 nm laser focused by a 1.2NA water immersion objective and 570-620 nm confocal detection by an avalanche photodiode. We use 80 nm apertures milled in similar conditions to the devices used for the UV experiments. Therefore the number of detected molecules and the size of the detection volume can be compared considering the difference in the illumination wavelength. Figure S11 shows FCS correlation functions recorded with various Alexa Fluor 546 concentrations. The FCS correlation data is fitted using a three dimensional Brownian diffusion model with an additional blinking term:[10]

$$G(\tau) = \frac{1}{N}\left[1 + \frac{T}{1-T}\exp\left(-\frac{\tau}{\tau_T}\right)\right]\left(1 + \frac{\tau}{\tau_d}\right)^{-1}\left(1 + \frac{1}{\kappa^2}\frac{\tau}{\tau_d}\right)^{-0.5} \tag{S1}$$

where N is the total number of molecules, $T$ the fraction of dyes in the dark state, $\tau_T$ the dark state blinking time, $\tau_d$ the mean diffusion time and $\kappa$ the aspect ratio of the axial to transversal dimensions of the nanohole volume. While the ZMW geometry obviously does not fulfill the assumption of free 3D diffusion, the above model equation was found to empirically describe well the FCS data inside ZMWs using an aspect ratio set to $\kappa = 1$ as found previously [9,11]. The dark state contribution remains quite small and is only needed to account for the fast dynamics below 10 µs. Owing to the larger statistical noise, this type of fluctuation is not currently detectable in the UV.

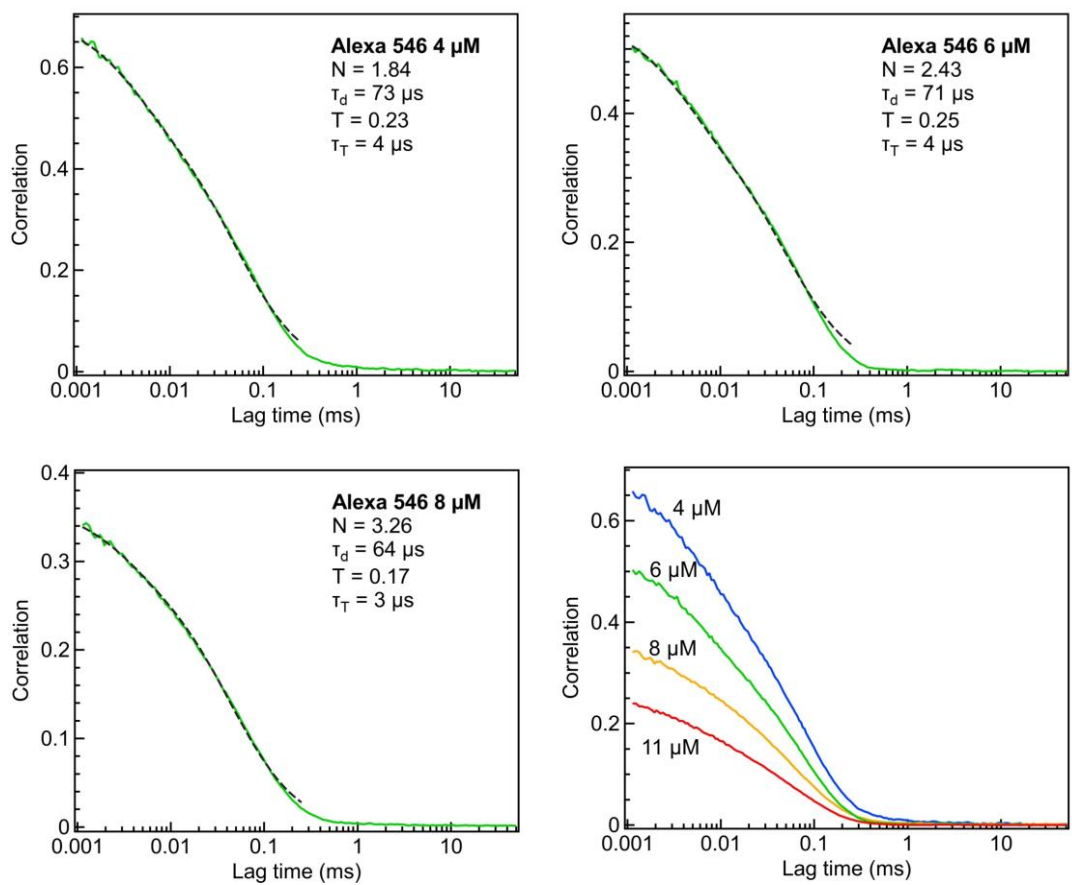

**Figure S11.** FCS correlation (green) and numerical fits (black dashed lines). The fit parameters are indicated each panel. The 80 nm ZMW is milled with identical parameters as the one used for the UV experiments.

**S12. Numerical simulations of the detection volume**

We use the wave optics module of COMSOL Multiphysics 5.5 to simulate the propagation of light inside a 80 nm diameter nanoaperture. The excitation is a plane wave with 266 or 557 nm. The vertical profile of the nanoaperture takes into account the tapering due to FIB milling and the 12 nm thick silica layer deposited by PECVD. To mimic our experiments, light is incoming from the bottom of the aperture where the diameter is the smallest. The inside volume of the aperture and the upper medium are set to a refractive index of water. A tetrahedral mesh is used with mesh size ranging from 0.3 nm to 10 nm. Scattering boundary conditions were used to suppress reflections from the domain boundaries.

Figure S12 show the simulation results for the two excitation wavelengths. The decay profile of the excitation intensity along the vertical center axis of the aperture is also represented. The light with the longer 557 nm wavelength has a shorter penetration inside the sub-wavelength aperture, explaining why the detection volume is smaller with 557 nm as compared to 266 nm excitation (Fig. 2g,h). To compute the FCS detection volume at each wavelength, we take into account the undercut into the quartz substrates, which add a constant volume of 0.45 aL for both wavelengths. Then we sum the volume from the undercut to the volume from the aperture, assuming for simplicity a monoexponential decay inside the aperture with characteristic decay length 27 nm at $\lambda$ = 557 nm and 35 nm at $\lambda$ = 266 nm. For the UV illumination, a cavity-like mode is excited, which shifts the attenuation decay by an extra 50 nm inside the aperture. With these values, we obtain a detection volume of 1.1 aL at 266 nm and 0.7 aL at 557 nm, in very good agreement with the experimental volumes (Fig. 2h) determined from the slope of the linear fits in Fig. 2g.

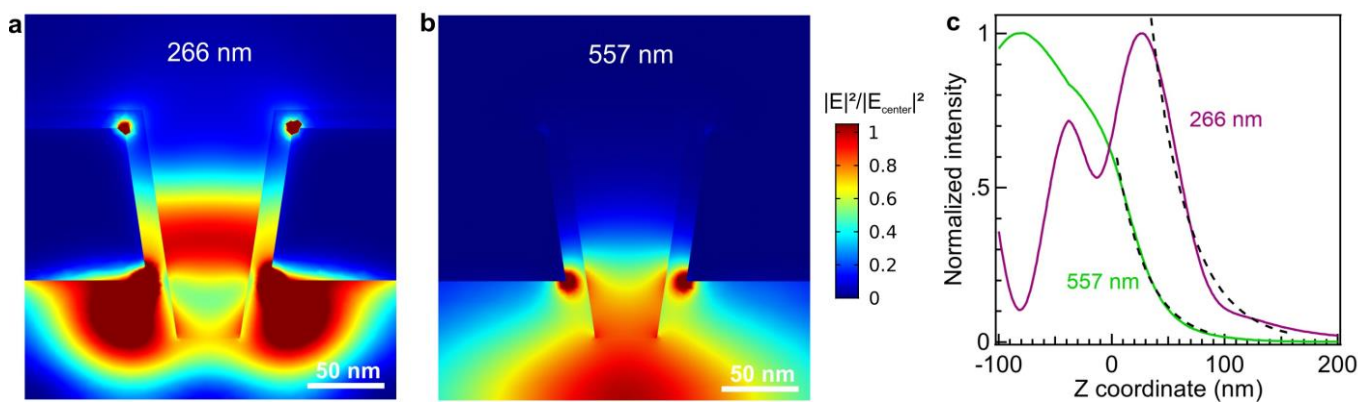

**Figure S12.** (a,b) Normalized intensity profiles at two different wavelengths 266 nm (a) and 557 nm (b) computed for a 80 nm diameter aperture milled in aluminum and covered by a 12 nm thick silica layer aiming at reproducing the experimental FIB-milled configuration (Fig. S3a). The peak enhancement value along the vertical center axis is 4.6 for 266 nm and 3.7 for 557 nm. (c) Comparison of the normalized decay profiles of the excitation intensities along the vertical center axis of the aperture. The origin (Z=0) is taken at the bottom quartz-aluminum interface. Monoexponential fits of the evanescently decaying sections are shown in black dashed lines.

## S13. Protein absorption and emission spectra

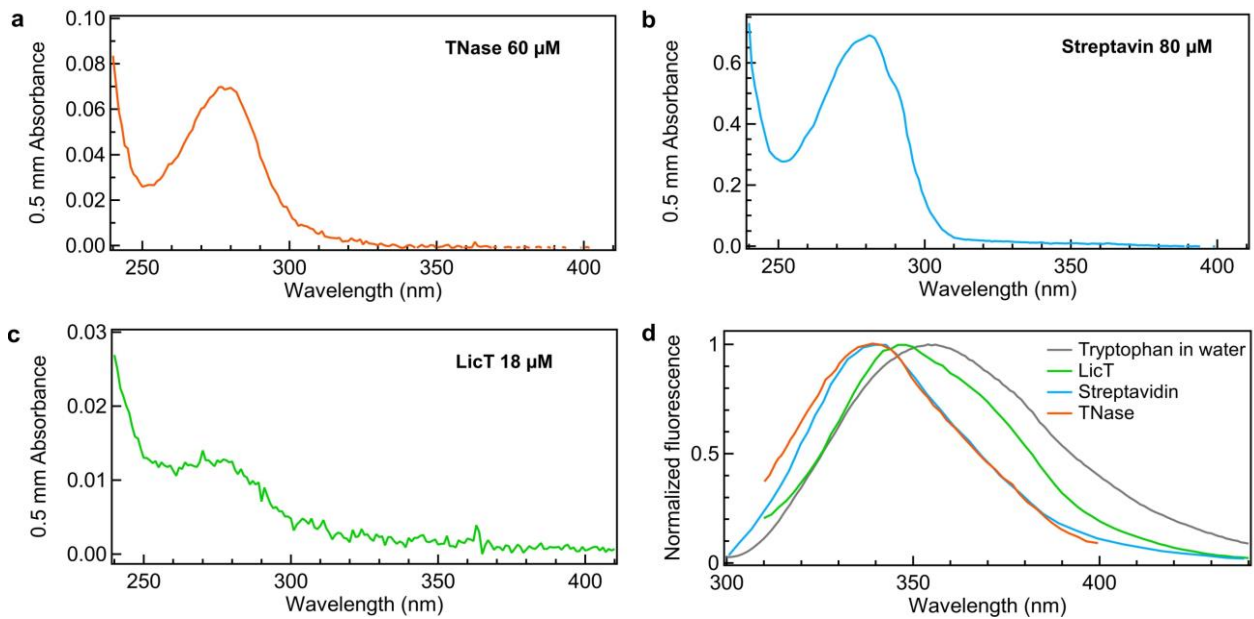

**Figure S13.** (a-c) Absorbance spectra corrected for background recorded for TNase, Streptavidin and LicT on a Tecan Spark 10M spectrofluorometer with 0.05 cm path length. (d) Normalized autofluorescence emission spectra for the proteins and pure tryptophan dissolved in water. The emission spectrum of tryptophan depends on its local environment, with a tendency to red-shift when exposed to water [12].

## 14. Signal to noise ratio in FCS in presence of background

The signal to noise ratio $SNR$ in a FCS experiment is commonly defined by $SNR = \frac{G(0)}{\sigma(G(0))}$ where $\sigma(G(0))$ is the standard deviation of the correlation amplitude $G(0)$. Following Koppel's seminal work about statistical accuracy in FCS,[13] the $SNR$ is generally derived from Eq. (40) in Ref.[13] as $SNR = CRM\sqrt{T_{tot}\,\Delta\tau}$, where $CRM$ denotes the fluorescence brightness (detected photons per second and per molecule), $T_{tot}$ the total integration time and $\Delta\tau$ the minimum lag time defining the time interval for computing the correlation. Note that Koppel's Eq. (40) contains a denominator taking the form $(1 + 4\,CRM\,\Delta\tau + 2\,CRM^2\Delta\tau\,\tau_d)^{1/2}$ where $\tau_d$ is the diffusion time. However in practice the correction introduced by this denominator is small and can generally be neglected. Note also that an extra term $\frac{1}{\sqrt{1+1/N}}$ can be introduced to account for possibly low average number of molecules in the detection volume [14] (Koppel's formula assumes $N \gg 1$).

It is important to stress out that the generally used formula $SNR = CRM\sqrt{T_{tot}\,\Delta\tau}$ does not consider the presence of background.[13] In presence of an extra background intensity $B$ (counts per second) on the detector, the $SNR$ expression must be modified. Combining Koppel's Eq.(40), (63) and (64) in ref.[13] and using our notations, the signal to noise ratio in FCS in presence of background can be expressed as:

$$SNR = \frac{CRM\left(\frac{SBR}{1+SBR}\right)\sqrt{T_{tot}\,\Delta\tau}}{(1 + 4\,CRM\,\Delta\tau\left(\frac{SBR}{1+SBR}\right) + 2\,CRM^2\left(\frac{SBR}{1+SBR}\right)^2\Delta\tau\,\tau_d)^{1/2}} \tag{S2}$$

In presence of background, this equation includes an extra term $\left(\frac{SBR}{1+SBR}\right)$ where $SBR = (N * CRM)/B$ is the signal to background ratio. In practice, the denominator in Eq. (S2) can often be neglected so that the simplified expression for the $SNR$ with background becomes

$$SNR = CRM\left(\frac{SBR}{1+SBR}\right)\sqrt{T_{tot}\,\Delta\tau} = CRM\left(1 - \frac{B}{B + N * CRM}\right)\sqrt{T_{tot}\,\Delta\tau} \tag{S3}$$

For large signal to background ratios, one retrieves Koppel's established formula, while for very low $SBR$ values, the signal to background becomes a linear prefactor lowering the $SNR$.

In our UV-FCS experiments with the horn antennas, our typical values are $CRM$ = 70 cts/s, $B$ = 600 cts/s, $N$ = 3 to 14 molecules, $T_{tot}$ = 90 s, $\Delta\tau$ = 10 µs. This gives signal to background ratios between 0.35 and 1.6, and signal to noise ratios between 0.5 and 1.3.

In the confocal case, using $CRM$ = 8 cts/s, $B$ = 2000 cts/s, $N$ = 10 molecules, $T_{tot}$ = 90 s, $\Delta\tau$ = 10 µs, the signal to background becomes 0.04 and the signal to noise is less than 0.01, preventing any possible experiment. Increasing the integration time to 1 hour and the minimum lag time to 100 µs, improves the $SNR$ to 0.18, yet this low value shows that any confocal UV-FCS experiment on single Trp protein would remain highly challenging.

Lastly, we would like to point out that maximizing the $SNR$ should not be the only consideration. As already pointed out by Koppel, another constraint is that $G(0)$ must remain much greater that the residual background correlation amplitude $\rho_B$: *"If this isn't much greater than [$\rho_B$], the experiment is in trouble"* (ref.[13] page 1944). In our antenna experiments, we have $\rho_B = 0.01$, while our different signal correlation amplitudes range from 0.022 to 0.04 (Fig. S7-S8), satisfying the condition $G(0) > \rho_B$.

The $SNR$ formula can be used to predict the feasibility of a UV-FCS experiment and provide guidelines so as to set the experimental conditions. Considering the case of free tryptophan molecules in solution (quantum yield 12%, expected brightness in the antenna 30 counts/s), the small molecular mass of 0.2 kDa implies a diffusion time below 10 μs imposing to set Δτ in the microsecond range. In the current conditions with 20 molecules in the antenna, the predicted $SNR$ for free tryptophan is 0.2, which remains too low to yield relevant FCS data. However, if with further work the background can be further reduced to 200 counts/s, and if the diffusion can be slowed down without introducing some photopolymerization artefacts, then UV-FCS on free tryptophans would become within experimental range.

## S15. Supplementary Methods

*Optical horn antennas*

The fabrication of horn antennas builds on our recent protocol [15]. We first mill the microreflector into a NEGS1 quartz coverslip covered with 100 nm aluminum using FIB (FEI dual beam DB235 Strata, 30 kV acceleration voltage, 300 pA ion current). The cone angle is chosen around 30° following our previous work [15]. A second 100 nm thick aluminum layer is deposited on top of the sample by electron-beam evaporation (Bühler Syrus Pro 710). Then a 80 nm diameter aperture is milled by FIB (10 pA current) in the center of the horn antenna unit with a 60 nm undercut chosen to optimize the signal to background ratio (Fig. S3) [16]. To protect the aluminum surface against corrosion [8,17,18], the horn antennas are covered by a 12 nm-thick conformal silica layer using plasma-enhanced chemical vapor protection (PECVD, PlasmaPro NGP80 from Oxford Instruments).

*Protein samples*

Thermonuclease staphylococcal nuclease from *Staphylococcus aureus* and streptavidin from *Streptomyces avidinii* are purchased from Sigma-Aldrich. Transcription antiterminator protein LicT from *Bacillus subtilis* is provided by Emmanuel Margeat, Nathalie Declerck and Caroline Clerté (CBS Montpellier, France) [19,20]. All details about the proteins used in this work are summarized in the Supporting Information Tab. S1. The proteins are dissolved in a Hepes buffer (25 mM Hepes, 300 mM NaCl, 0.1 v/v% Tween20, 1 mM DTT, and 1 mM EDTA 1 mM at pH 6.9). The protein solutions are centrifuged for 12 min at 142,000 g using an air centrifuge (Airfuge 20 psi) and the supernatants are stored in small aliquots at -20°C. The concentrations are assessed using a spectrofluorometer (Tecan Spark 10 M, Fig. S13) using the extinction coefficients derived from the protein sequence and summarized in Tab. S1. 10 mM of mercaptoethylamine MEA and 25 mM of glutathione GSH (both from Sigma Aldrich) are added to the buffer just before the experiments to improve the photostability and neutralize the reactive oxygen species (Fig. S5). Oxygen removal using degassing did not modify significantly the TNase autofluorescence signal. Therefore, we decided to not undertake any special action to remove the oxygen dissolved in the buffer solution for the TNase and the streptavidin experiments. For LicT experiments, the oxygen dissolved in the solution is removed by bubbling the buffer with argon for at least 5 minutes, as it has a significant impact on the LicT autofluorescence signal (Fig. S5f). The solution is then quickly placed on the UV microscope and the chamber is filled with argon and covered with a coverslip to prevent the oxygen from the air to enter into the solution. The LicT autofluorescence remained stable for about one hour, showing no sign of oxygen rediffusion during this period.

*Experimental setup*

The laser source is a 266 nm picosecond laser (Picoquant LDH-P-FA-266, 70 ps pulse duration, 80 MHz repetition rate) with an average power of 10 μW. The laser beam is spatially filtered by a 50 μm pinhole to ensure a quasi-Gaussian profile. A dichroic mirror (Semrock FF310-Di01-25-D) reflects the laser beam towards the microscope. The UV objective is a LOMO 58x 0.8 NA with water immersion. The optical horn antenna is positioned at the laser focus with a 3-axis piezoelectric stage (Physik Instrumente P-517.3CD). The fluorescence light is collected by the same microscope objective and focused onto a 80 μm pinhole by a quartz lens with 200 mm focal length (Thorlabs ACA254-200-UV). A stack of three emission filters (Semrock FF01-300/LP-25, FF01-375/110-25 and FF01-334/40-25) selects the fluorescence photons within the 310-360 nm range. The detection is performed by a photomultiplier tube (Picoquant PMA 175) connected to a time-correlated single photon counting TCSPC module (Picoquant Picoharp 300 with time-tagged time-resolved mode). The integration time is 90 sec per antenna. Since we work at quasi-neutral pH, some photocorrosion of the aluminum can still occur after a longer UV exposure [17], this is why we decide to limit the illumination time to ensure the maximum data reproducibility. For each horn antenna used in this work, we record the background intensity $B$ by replacing the protein solution by the buffer only, all the other experimental conditions are kept identical.

*Fluorescence correlation spectroscopy*

The fluorescence time traces data are analyzed with Symphotime 64 (Picoquant) and Igor Pro 7 (Wavemetrics). For the time gating, only the photons within a 3 ns window after the fluorescence peak are selected for the analysis. The rationale behind this choice is that for an exponential decay of characteristic time τ, 95% of the signal is collected within a time window of width 3τ. We thus consider a time window corresponding to 3× the fluorescence lifetime (which is about 1 ns for tryptophan in the horn antenna for the different proteins tested here) in order to provide a good trade-off between signal collection and noise rejection.

The FCS correlations are computed using Symphotime 64 and fitted with a two species model:[21]

$$G(\tau) = \rho_1 \left(1 + \frac{\tau}{\tau_1}\right)^{-1} \left(1 + \frac{1}{\kappa^2}\frac{\tau}{\tau_1}\right)^{-0.5} + \rho_2 \left(1 + \frac{\tau}{\tau_2}\right)^{-1} \left(1 + \frac{1}{\kappa^2}\frac{\tau}{\tau_2}\right)^{-0.5} \quad (S4)$$

where $\rho_i$ and $\tau_i$ are the amplitude and diffusion time of each species and κ is the aspect ratio of the axial to transversal dimensions of the detection volume (set to κ = 1 for the horn antenna following our previous works [15,16,22]). The rationale behind this two species model is that the first fast-diffusing species accounts for the protein while the second slow-diffusing species corresponds to the residual background. In the absence of any protein sample (using only the same buffer solution), a correlation is still observed from the background with an amplitude below 0.01 and a characteristic time of 50 ms (Fig. S7-S9). The origin of this background correlation remains unclear, the 50 ms time (20 Hz) suggests some remaining mechanical vibration or electric noise on our microscope. As the diffusion times of the proteins are below 500 μs, the background contribution ($\rho_2$, $\tau_2$) can be readily separated on the FCS data. All the fit results are detailed in the Supporting Information Fig. S7 & S8. From the correlation amplitude $\rho_1$, the total fluorescence intensity $F$ and the background intensity $B$, we compute the background-corrected number of proteins [21] as $N = (1 - B/F)^2 /\rho_1$ and the fluorescence brightness per protein as $CRM = (F - B)/N$ [16]. As we consider lag times longer than 10 μs for the UV-FCS analysis, the afterpulsing from the photomultiplier and the repetition rate of the laser are not a problem.

**Supplementary references**